\documentclass[fleqn,usenatbib,letterpaper]{mn2e}
\usepackage{graphicx}
\usepackage{amsmath}
\def\jnl@style{\it}
\def\aaref@jnl#1{{\jnl@style#1}}
\def\aaref@jnl#1{{\jnl@style#1}}

\def\aj{\aaref@jnl{AJ}}                   % Astronomical Journal
\def\araa{\aaref@jnl{ARA\&A}}             % Annual Review of Astron and Astrophys
\def\apj{\aaref@jnl{ApJ}}                 % Astrophysical Journal
\def\apjl{\aaref@jnl{ApJ}}                % Astrophysical Journal, Letters
\def\apjs{\aaref@jnl{ApJS}}               % Astrophysical Journal, Supplement
\def\ao{\aaref@jnl{Appl.~Opt.}}           % Applied Optics
\def\apss{\aaref@jnl{Ap\&SS}}             % Astrophysics and Space Science
\def\aap{\aaref@jnl{A\&A}}                % Astronomy and Astrophysics
\def\aapr{\aaref@jnl{A\&A~Rev.}}          % Astronomy and Astrophysics Reviews
\def\aaps{\aaref@jnl{A\&AS}}              % Astronomy and Astrophysics, Supplement
\def\azh{\aaref@jnl{AZh}}                 % Astronomicheskii Zhurnal
\def\baas{\aaref@jnl{BAAS}}               % Bulletin of the AAS
\def\jrasc{\aaref@jnl{JRASC}}             % Journal of the RAS of Canada
\def\memras{\aaref@jnl{MmRAS}}            % Memoirs of the RAS
\def\mnras{\aaref@jnl{MNRAS}}             % Monthly Notices of the RAS
\def\pra{\aaref@jnl{Phys.~Rev.~A}}        % Physical Review A: General Physics
\def\prb{\aaref@jnl{Phys.~Rev.~B}}        % Physical Review B: Solid State
\def\prc{\aaref@jnl{Phys.~Rev.~C}}        % Physical Review C
\def\prd{\aaref@jnl{Phys.~Rev.~D}}        % Physical Review D
\def\pre{\aaref@jnl{Phys.~Rev.~E}}        % Physical Review E
\def\prl{\aaref@jnl{Phys.~Rev.~Lett.}}    % Physical Review Letters
\def\pasp{\aaref@jnl{PASP}}               % Publications of the ASP
\def\pasj{\aaref@jnl{PASJ}}               % Publications of the ASJ
\def\qjras{\aaref@jnl{QJRAS}}             % Quarterly Journal of the RAS
\def\skytel{\aaref@jnl{S\&T}}             % Sky and Telescope
\def\solphys{\aaref@jnl{Sol.~Phys.}}      % Solar Physics
\def\sovast{\aaref@jnl{Soviet~Ast.}}      % Soviet Astronomy
\def\ssr{\aaref@jnl{Space~Sci.~Rev.}}     % Space Science Reviews
\def\zap{\aaref@jnl{ZAp}}                 % Zeitschrift fuer Astrophysik
\def\nat{\aaref@jnl{Nature}}              % Nature
\def\iaucirc{\aaref@jnl{IAU~Circ.}}       % IAU Cirulars
\def\aplett{\aaref@jnl{Astrophys.~Lett.}} % Astrophysics Letters
\def\apspr{\aaref@jnl{Astrophys.~Space~Phys.~Res.}}
\def\bain{\aaref@jnl{Bull.~Astron.~Inst.~Netherlands}} 
\def\fcp{\aaref@jnl{Fund.~Cosmic~Phys.}}  % Fundamental Cosmic Physics
\def\gca{\aaref@jnl{Geochim.~Cosmochim.~Acta}}   % Geochimica Cosmochimica Acta
\def\grl{\aaref@jnl{Geophys.~Res.~Lett.}} % Geophysics Research Letters
\def\jcp{\aaref@jnl{J.~Chem.~Phys.}}      % Journal of Chemical Physics
\def\jgr{\aaref@jnl{J.~Geophys.~Res.}}    % Journal of Geophysics Research
\def\jqsrt{\aaref@jnl{J.~Quant.~Spec.~Radiat.~Transf.}}
\def\memsai{\aaref@jnl{Mem.~Soc.~Astron.~Italiana}}
\def\nphysa{\aaref@jnl{Nucl.~Phys.~A}}   % Nuclear Physics A
\def\physrep{\aaref@jnl{Phys.~Rep.}}   % Physics Reports
\def\physscr{\aaref@jnl{Phys.~Scr}}   % Physica Scripta
\def\planss{\aaref@jnl{Planet.~Space~Sci.}}   % Planetary Space Science
\def\procspie{\aaref@jnl{Proc.~SPIE}}   % Proceedings of the SPIE

\begin{document}
\title[Cores in Classical dSphs?]{Cores in Classical Dwarf Spheroidal Galaxies? A Dispersion-Kurtosis Jeans Analysis Without Restricted Anisotropy}% 
\author[T. Richardson \& M. Fairbairn]{Thomas Richardson\thanks{thomas.d.richardson@kcl.ac.uk}$^{1}$, Malcolm Fairbairn\thanks{malcolm.fairbairn@kcl.ac.uk}$^{1}$\\$^{1}$Physics, Kings College London, Strand, London WC2R 2LS, UK}
\maketitle
\begin{abstract}
We attempt to measure the density of dark matter in the two Dwarf Spheroidal Galaxies Fornax and Sculptor using a new method which employs Jeans equations based on both the second and fourth moment of the Collisionless Boltzmann Equation (i.e. variance {\it and} kurtosis of line of sight stellar velocities).  Unlike previous related efforts, we allow the anisotropy of the radial and tangential second and fourth order moments to vary independently of each other.  We apply the method to simulated data and establish that to some degree it appears to be able to break the degeneracies associated with second order only Jeans analyses.  When we apply the technique to real data, we see no huge improvement in our understanding of the inner density of Fornax, which can still be fit by either a quite cuspy or cored density profile.  For Sculptor however we find that the technique suggests that the data is incompatible with a steep profile and a cored profile seems more consistent.  As well as presenting these new results and comparing them to other estimates in the literature, we try to understand why the technique is more effective in one system than the other.
\end{abstract}


\section{Introduction}
It is well documented that the $\Lambda$CDM model of cosmology is remarkably successful at reproducing the observed large scale structure of the universe \citep{SDSS} in dissipationless N-body simulations of cold dark matter \citep[e.g][]{NFWls}. At galactic scales however there are a number of discrepancies between the simulation predictions \citep{NFWhalos,aquarious,bolshoisim} and observations including (but not limited to) the abundance \citep{misssats,wherearethey} and phase space correlation \citep{mwpancake,satphase} of satellite galaxies, the absence of highly luminous satellite galaxies \citep{toobigtofail} and the observation of seemingly cored density profiles at the centre of low surface brightness spiral \citep{spiralcores} and dwarf spheroidal galaxies \citep{penarrubia, amoriscomulti} that is at odds with the more cusped density profiles ubiquitous in the halos of cold dark matter simulations. The interaction of dark matter with baryons, absent in most simulations, could ameliorate \citep{astrocore,nomorecusp,astrocorefeedback} any of these differences to a greater or lesser extent and is an area of active study. Work has also been undertaken to investigate the possibility that the Milky Way is a statistical outlier \citep{milkywayrare, noplacelikehome}.    

The interesting anomalies have also given rise to a number of alternative cosmological models. Warm dark matter is one such contender with a free-streaming length that preserves the successful large-scale formation \citep{warmoklarge} whilst potentially reducing the abundance \citep{warmlesssats} and size \citep{warmsize} of satellites in line with observation. It has also been suggested \citep{walkSI} that relaxing the assumption of collisionless dark matter \citep{firstSI} could generate extended density cores.    

To minimise baryonic effects, the high mass-luminosity ratio observed in dwarf spheroidal galaxies makes them a relatively 'clean' target to test predictions from $\Lambda$CDM simulations and herein we use satellites of the Milky Way to try and tackle the cusp vs core issue. This property also makes them good candidates for indirect detection searches of dark matter such as that with the Fermi telescope \citep{fermidwarf} and a precise measurement of the density profile is key to reducing the large uncertainties in the the expected flux of incident annihilation products such as gamma rays.

 Several methods have been proposed to infer the density of dwarf spheroidal galaxies from a limited sample of projected radii and velocity measurements. A key difficulty that the majority of these methods face is that to infer the gravitational potential one must contend with the six dimensional phase space distribution function $f(\textbf{r},\textbf{v})$ as an unknown nuisance parameter which is further obscured from the observed velocity data by projection along the line of sight.

 Perhaps the simplest method is to use the Jeans equations \citep{binney} that relate the underlying gravity potential to the more manageable moments of the phase space distribution function. As partial solutions to the collisionless Boltzmann equation however then with a truncated series one cannot guarantee that all solutions correspond to physical distribution functions and must contend with degeneracies \citep{merritt87} upon projection along the line of sight. The degeneracies associated with the traditional second order Jeans analysis are more pronounced for a flat stellar velocity dispersion profile. Being an analytic method it is also necessary to parametrise the anisotropy and density that for mathematical simplicity alone have led to constraining and often unphysical additional assumptions such as Gaussianity and constant anisotropy in the literature. 

To alleviate issues with the Jeans equations raised above, numerical procedures such as the orbit-superposition method \citep{Schwarzschild} utilise the Jeans theorem to ensure physical solutions to the collisionless Boltzmann equation and without any assumed form for the anisotropy. Such methods have not yet however been able to distinguish cusps from cores in classical dwarf spheroidals \citep{Breddels, Jardel}. By considering the simpler analytic Jeans analysis, \cite{Wolf} discovered that the mass at the half-light radius is invariant under a general choice of anisotropy. With the additional discovery of multiple stellar populations in Fornax and Sculptor dwarf spheroidals this was used to devise a method \citep{penarrubia} that places perhaps the strongest constraints in favour of cored density profiles in dwarf spheroidals.

 To test the assumptions mentioned in the paragraph above that underpin this result and to alleviate the degeneracy that blights a dispersion only method we were motivated to extend the truncated Jeans analysis to fourth order. In the context of dwarf spheroidals this was first proposed in \cite{Lokas02} and was applied to Draco \cite{Lokas05} but was limited to a constant anisotropy parameter and phase space distribution functions with correlated second and fourth moments where tangentially biased variances necessarily implied tangentially biased kurtosis. In previous work \cite{me} we provided a framework that, by introducing anisotropy parameters analogous to Binney's at second order, imposed no assumptions on the form or correlations of anisotropy parameters. This extension facilitates the most generalised representation of anisotropy yet used in an analytic Jeans analysis that complements numeric approaches. 

A brief summary of the fourth order Jeans analysis and the statistical likelihood analysis used to fit this to the velocity data is provided in Sections 2 and 3  and we refer the reader to \cite{me} for a detailed account. In Section 4 we display the results of the analysis on data sets from Fornax and Sculptor dwarf spheroidal galaxies. Final we discuss the implications of the results in comparison to other works and make concluding remarks.              
   
\section{Solutions to the Jeans Equations at Fourth Order}

One particular means to infer the DM dominated density of a dwarf spheroidal galaxy from a set of stellar positions and velocities is the collsionless Boltzmann equation \citep{merry} which for equilibrated spherically symmetric systems is 
\begin{center}
\begin{eqnarray}\label{boltz}
\frac{\partial f}{\partial t} &=& v_r \frac{\partial f}{\partial r} + \left( \frac{v^{2}_{\theta}+v^{2}_{\phi}}{r}-\frac{d\Phi}{dr}\right) \frac{\partial f}{\partial v_r} \nonumber\\ &+& \frac{1}{r}(v^{2}_{\phi}\cot \theta - v_rv_{\theta})\frac{\partial f}{\partial v_{\theta}}  \\ &-&  \frac{1}{r}(v_{\phi}v_r+v_{\phi}v_{\theta}\cot \theta )\frac{\partial f}{\partial v_{\phi}} \nonumber\\ &=& 0. \nonumber
 \end{eqnarray}
\end{center}
where $f(r,\textbf{v})$ is the four dimensional distribution function of stellar radii and velocity components in spherical coordinates $v_r, v_{\theta}, v_{\phi}$ and $\Phi(r)$ is the spherically symmetric gravitational potential that we assume to be dominated by the dark mass component. 
Equation \eqref{boltz} alone is not easy to work with but by multiplying through with a suitable factor and integrating over all velocities \citep{merry} one obtains the Jeans equations that stipulate conditions on the moments of the distribution for equilibrated systems,
\begin{equation}\label{moms}
\nu \overline{v^{2i}_r v^{2j}_{\theta} v^{2k}_{\phi}} = \int v^{2i}_r v^{2j}_{\theta} v^{2k}_{\phi} f(r,\textbf{v}) d^3v.
\end{equation}
of which $\nu(r)$, the zeroth moment that effectively normalises the distribution function is the observed local stellar density and the others may be written \citep{dejonghe86} as 
\begin{eqnarray}
\overline{v^{2i}_r v^{2j}_{\theta} v^{2k}_{\phi}} = \frac{1}{\pi} B(j+\frac{1}{2},k+\frac{1}{2})\overline{v^{2i}_r\; v^{2(j+k)}_t}
\end{eqnarray}
where $B(x,y)$ is the Beta function and $v_t = \sqrt{v^{2}_{\theta}+v^{2}_{\phi}}$ is the magnitude of the two dimensional tangential velocity. Due to the assumption of spherical symmetry net rotations and all other odd moments vanish.
 The two moments at second order, i.e the variances $\sigma^2_r \equiv \overline{v^2_r},\sigma^2_t \equiv \overline{v^2_t}$ are related to the underlying gravitational potential by the well-known second order Jeans equation \citep{binney},
\begin{equation}\label{jeans}
\frac{d(\nu \sigma^{2}_{r})}{dr} + \frac{2}{r}\nu(2 \sigma^{2}_{r}-\sigma^2_t) +\nu \frac{d \Phi}{dr} = 0.
\end{equation}   
At fourth order there are two unique Jeans equations \citep{merry} that relate the three fourth order moments,
\begin{equation}
\label{hojeans1}
\frac{d(\nu \overline{v^{4}_{r}})}{dr} - \frac{3}{r}\nu \overline{v^{2}_{r}v^{2}_{t}}  + \frac{2}{r}\nu \overline{v^{4}_{r}}  + 3 \nu \sigma^{2}_{r} \frac{d \Phi}{dr} = 0
\end{equation}
\begin{equation}
\label{hojeans2}
\frac{d(\nu \overline{v^{2}_{r}v^{2}_{t}})}{dr} - \frac{1}{r}\nu \overline{v^{4}_{t}}  + \frac{4}{r}\nu \overline{v^{2}_{r}v^{2}_{t}}  +  \nu \sigma^{2}_{t} \frac{d \Phi}{dr} = 0.
\end{equation}
With $n$ equations and $n+1$ moments at the $2n^{th}$ order then even if one has knowledge of the stellar and DM densities there still persists a degeneracy of solutions to the Jeans equations at each order. Upon projection along the line of sight this unknowable choice gives rise to the degeneracy problem. To represent the full set of degenerate solutions the parameter, 
 \begin{equation}
\label{anis}
\beta(r) \equiv 1-\frac{\sigma^{2}_{t}(r)}{2\sigma^{2}_{r}(r)},
\end{equation}
which measures the second order deviation of the system from isotropy \footnote{i.e the difference in widths of the radial and angular velocity distributions}, was introduced \citep{binney} to fully constrain \eqref{jeans} and its specification defines the solution.
We replicate this idea at fourth order by introducing an analog to Binney's anisotropy parameter
\begin{equation}\label{betprime}
\beta'(r) = 1-\frac{3}{2}\frac{\overline{v^{2}_{r}v^{2}_{t}}}{\overline{v^{4}_{r}}},
\end{equation}
that represents the deviation of the fourth order moments from the isotropic system ${\overline{v^{4}_{r}}} = \frac{3}{2}\overline{v^{2}_{r}v^{2}_{t}}$ and therefore naively \citep[see][for details]{me} determines the difference between the kurtosis of the radial and angular velocity distributions. All solutions to the Jeans equations at fourth order may then be represented by the ordinary differential equations,
\begin{equation}\label{lojeans}
\frac{d(\nu \sigma^{2}_{r})}{dr} + \frac{2\beta}{r}\nu \sigma^{2}_{r} +\nu \frac{d \Phi}{dr} = 0.
\end{equation}
\begin{equation}\label{radeq}
\frac{d(\nu \overline{v^{4}_{r}})}{dr} + \frac{2 \beta'}{r}\nu \overline{v^{4}_{r}} + 3 \nu \sigma^{2}_{r} \frac{d \Phi}{dr} = 0,
\end{equation}
from which one can deduce the remaining moments with $\sigma^2_t = 2(1-\beta)\sigma^2_r$, $\overline{v^{2}_{r}v^{2}_{t}}=\frac{2}{3}(1-\beta^{\prime})\overline{v^4_r}$ and
\begin{equation}\label{tanfourgen}
\overline{v^4_t} = \frac{4}{3}\left((1-\beta')(2-\beta') - \frac{r}{2} \frac{d\beta'}{dr}\right) \overline{v^4_r}+2(\beta'-\beta)r\sigma^{2}_{r}\frac{d\Phi}{dr}.\end{equation}
If one assumes the anisotropy to be constant with radius then an integral form exists \citep{Lokas02} that simplifies the calculation,
 \begin{equation}
\nu \sigma^{2}_{r}(\beta=\rm{const})=r^{-2\beta}\int^{\infty}_{r}r^{2\beta}\nu\frac{d\Phi}{dr}dr
\end{equation} 
\begin{equation}
\nu \overline{v^{4}_{r}}(\beta^{\prime} =\rm{const})=3r^{-2\beta^{\prime}}\int^{\infty}_{r}r^{2\beta^{\prime}}\nu\sigma^{2}_{r}\frac{d\Phi}{dr}dr
\end{equation} 
although in this work we will be considering situations where both anisotropy parameters vary with radius.
Once we are in possession of all the necessary moments it is possible to project along the line of sight and calculate the moments of the line-of-sight velocity distribution,
\begin{equation}\label{LOSsecond}
\Sigma \sigma^{2}_{p}(R) = 2\int^{\infty}_{R} (1-\beta\frac{R^2}{r^2}) \frac{\nu \sigma^{2}_{r} r}{\sqrt{r^2-R^2}}dr 
\end{equation}
\begin{equation}\label{LOSgenfour}
\Sigma\overline{v^{4}_{p}}(R) =  2\int^{\infty}_{R} \left(g(\beta')\overline{v^4_{r}}+\frac{3R^4}{4r^3}(\beta'-\beta)\sigma^{2}_{r}\frac{d\Phi}{dr}\right) \frac{\nu(r)r}{\sqrt{r^2-R^2}}dr
\end{equation}
where $R$ is the observed, projected radius perpendicular to the line of sight.  The function $g(\beta',r,R)$ common to the separable augmented density system is 
\begin{equation}\label{gfunc}
 g(\beta',r,R)=1-2\beta'\frac{R^2}{r^2}+\frac{\beta'(1-\beta')}{2}\frac{R^4}{r^4}-\frac{R^4}{4r^3}\frac{d\beta'}{dr} 
\end{equation}
and we note the general Jeans solution extends the work of \cite{Lokas02} for which $\beta'=\beta={\rm constant}$ and its generalisation to arbitrary anisotropy, the separable augmented density model \citep[see for e.g.][]{an11a} defined by $\beta'(r)=\beta(r)$, with an additional term.   
These projected moments provide a means of comparison with observable data alongside the surface density
\begin{equation}
\Sigma(R) = 2\int^{\infty}_{R} \frac{\nu r}{\sqrt{r^2-R^2}}dr.
\end{equation}

Now we are in possession with expressions for the second and fourth moments of the line of sight velocity dispersion for a given underlying dark matter density, we need to see how well we can constrain the unknown dark matter density using data.

\section{Likelihood Analysis}
In this section we briefly summarise the Jeans/Monte-Carlo Markov chain (MCMC) methodology presented in \cite{me} by which the posterior distributions for a set of anisotropy and density parameters $p$ are generated from a data set $d$ of stellar line-of-sight radii and velocities via likelihood function $\mathcal{L}(d|p)$.
\subsection{Anisotropy and Density Parameterisations}

With the ultimate aim of quantifying the uncertainty in a measurement of a dwarf spheroidal's density profile a parametric form for the anisotropy and density is required for an MCMC analysis. In particular we are motivated to consider realistic density profiles which can exhibit either extended cores or the cuspy centres predicted by dissipationless DM only simulations. To this end the Einasto parametrisation \citep{einasto} of the density and its corresponding logarithmic density slope are
\begin{equation}
\rho_{\rm{dm}}(r) = \rho_{-2}\exp\left\{-\frac{2}{\alpha}\left[\left(\frac{r}{r_{-2}}\right)^{\alpha}-1\right]\right\}
\end{equation}
\begin{equation}
\gamma(r) \equiv \frac{d\ln \rho}{d\ln r} = -2\left(\frac{r}{r_{-2}}\right)^{\alpha}
\end{equation}
where $r_{-2}$ is a scale radius that indicates where the logarithmic density slope $\gamma(r)= -2$, $\rho_{-2}$ is the scale density at this radius and $\alpha$ is a constant shape parameter that determines the rate at which the density slope deviates from -2 at the scale radius. Though all Einasto profiles are cored at the galactic centre $\gamma(r\to0)=0$, variation in the shape parameter $\alpha$ can mimic cusped profiles at all but vanishingly small radii. If the shape parameter is small $\alpha<1$ then $\gamma$ varies slowly from the scale radius $r_{-2}$ to the centre of the galaxy, extending the steeper gradient and wiping out the core. An estimate of the core size can be defined as,
\begin{equation}\label{coreradius}
\log_{10} r_{\rm{c}} = \log_{10} r_{-2} - \frac{1}{\alpha}
\end{equation}
which indicates the radius at which from zero the slope falls to  $\gamma=-0.2$ and we note that increasing $\alpha$ shifts the core radius towards $r_{-2}$.     
 To model the anisotropy parameters we use \citep{baes07},
\begin{equation}\label{bvh}
\beta(r) = (\beta_{\infty}-\beta_0) \frac{r^2}{r^{2}_{\beta}+r^2} + \beta_0
\end{equation}
which generically describes anisotropy with a quadratic transition about $r_{\beta}$ that asymptotes at $\beta_0$ for $r=0$ and $\beta_{\infty}$ for $r\rightarrow \infty$. We will use the same functional form for the radial dependence of $\beta'(r)$, with all parameters corresponding to $\beta^{\prime}$ denoted with a prime.
At second order the parameter set is thus $p_2 = \{\beta_0,\beta_{\infty},r_{\beta},\rho_{-2},r_{-2},\alpha\}$ which is extended at fourth order by the additional fourth order anisotropy parameters, $p_4 = \{\beta_0,\beta_{\infty},r_{\beta},\beta_0',\beta_{\infty}',r_{\beta}', \rho_{-2},r_{-2},\alpha\}$.

The local density $\nu(r)$ is modelled by a Plummer profile,
\begin{equation}
\nu(r) \propto \left(1+\frac{r^2}{R^{2}_{1/2}}\right)^{-\frac{5}{2}}
\end{equation}
where $R_{1/2}$ is the Plummer radius which is equivalent to the projected half-light radius that encompasses half of the stars on the circular projected surface. This parameter is taken from observational data and we are motivated to use this simple approximation to reduce the already considerable number of free parameters in the analysis. 

We note that this is the most rigid part of the analysis and that imposing a central core for the stellar density can have a pronounced impact on the dispersion at small radii that limits the range of physical solutions (\cite{evanscentre}). Assuming that the dark matter and tracer density have central slopes shallower than the isothermal cusp $\rho(r) \propto r^{-2}$ at the galactic centre and also that the anisotropy parameter asymptotes to a flat constant value $\beta_0$ as modelled above, then the anisotropy term dominates the mass term such that,
\begin{equation}
\frac{d\sigma^{2}_{r}}{dr} \approx \frac{\gamma_\star -2\beta_0}{r}\sigma^{2}_{r}
\end{equation}
where $\gamma_\star = - d\ln \nu / d\ln r$ is the logarithmic density slope of the tracer stars. With these assumptions then, that encompass cored and NFW dark matter density profiles with logarithmic slopes of $\gamma = 0$ and  $\gamma = 1$ respectively the dispersion is,
\begin{equation}
\lim_{r \to 0} \sigma^{2}_{r} \propto r^{\gamma_\star-2\beta_0}.
 \end{equation}
The seemingly benign assumption of a cored Plummer profile and spherical symmetry therefore demands that the dispersions of tangentially biased systems $\beta_0 < 0$ vanish and whilst for radially biased systems $\beta_0 > 0$ they diverge at the centre. One must therefore take care when interpreting results near the galactic centre and clearly the assumption of constant anisotropy at all radii coupled with a fixed tracer density is liable to these assumptions. Therefore though we adopt a Plummer profile with a fixed density slope the freedom of $\beta_0$ in the generalised parametrisation of $\beta$ effectively absorbs $\gamma_\star$, minimising the impact of the plummer profile assumption to some extent .    
 
 \subsection{Sample Moments and Likelihood Function}
With the Jeans equations defined in Section 2 it is possible to obtain the dispersion $\sigma^{2}_{p}$ and kurtosis $\kappa_p$ of the line of sight velocity distribution corresponding to a set $p$ of anisotropy and density parameters. We then fit these to the data set $d$ of $N_\star$ stellar line of sight velocities $\{v_i\}$ and positions $\{R_i\}$ with a suitable likelihood $\mathcal{L}(d|p)$ which denotes the probability that $d$ is drawn from a phase space distribution function with moments $\sigma^{2}_{p}(R)$ and $\kappa_p(R)$. Whilst one maximises the information in the data set by defining the total likelihood as the product of individual tracer probabilities, $\mathcal{L}(d|p) = \prod^{N_\star}_{i} \mathcal{F}(v_i|\sigma^{2}_{p}(R_i),\kappa_p(R_i))$, it is difficult to explicitly encompass the highly significant statistical fluctuations from limited sampling into the probability distributions $\mathcal{F}$. As such we split the data into $N_r$ radial bins of equal stellar content $N = N_\star / N_r$ and use the sample moment estimators,
\begin{equation}\label{sampdisp}
\widehat{\sigma}^{2} = \frac{1}{N} \sum^{N}_{i=1} (v_i - \mu )^2
\end{equation}
\begin{equation}\label{sampkurt}
\widehat{\kappa} = \frac{1}{N} \sum^{N}_{i=1} \frac{(v_i - \mu )^4}{(\widehat{\sigma}^{2})^2}
\end{equation}
where $\mu$ is the mean velocity of the bin, to represent the velocity data. In this way, wherein a compromise is made between statistical precision and spatial resolution, the probability distribution $\mathcal{S}_x(\widehat{x}|\sigma^{2}_{p}(R_{\rm{bin}}), \kappa_p(R_{\rm{bin}}))$ of the estimator $\widehat{x}$ has a dependence on the sample size that can be calculated numerically by simply taking many bootstrap samples of size $N$ from a suitable choice of parent distribution $\mathcal{F}(v|\sigma^{2}_{p}(R_{\rm{bin}}),\kappa_p(R_{\rm{bin}}))$ and adding Gaussian noise to represent the experimental errors. The distribution of kurtosis estimators was found to be both significantly biased and skewed when realistic samples of $N<400$ velocity measurements were drawn from parent distributions for which the variance and kurtosis are known (we use both Pearson and Gaussian superposition as our non-Gaussian profiles). To better approximate Gaussianity the kurtosis estimator was transformed \citep{Lokas05} to $\kappa^{\prime} = (\ln \kappa)^{0.1}$ but a significant skewness persisted for leptokurtic parent distributions. For this reason a $\chi^2$ form for the likelihood is inappropriate and we instead took of order $10^6$ bootstrap samples to generate a smooth\footnote{Maximum likelihood estimators were also considered but whilst they reduced bias and statistical errors the sampling distributions also displayed a significant skew. Maximising the 2D parameter space proved too computationally expensive to generate sufficient samples for a smooth numerical distribution.}, fully numeric probability distribution by binning the sample kurtosis data. 

For simplicity the variance and kurtosis estimators were considered independent in our analysis \footnote{Though upon measurement the estimators are in fact weakly correlated with Pearson coefficient $\rho \approx 0.2$} and we found that the marginalised kurtosis distribution $\mathcal{S}_{\kappa^{\prime}}$ is almost completely independent of the input variance, so we assume it is entirely independent. With these assumptions we employ the following likelihood for the joint analysis of dispersion and kurtosis,
\begin{equation}\label{like}
\mathcal{L}(d|p) = \prod^{N_r}_{i} \mathcal{S}_\sigma(\widehat{\sigma}^2_i|\sigma^2_p(R_i),\kappa^{\prime}_p(R_i)) \times \mathcal{S}_\kappa(\widehat{\kappa}_i^{\prime}|\kappa^{\prime}_{p}(R_i)).
\end{equation} 
and for comparison we use the dispersion-only likelihood,
\begin{equation}\label{likesec}  
\mathcal{L}_\sigma(d|p) = \prod^{N_r}_{i} \mathcal{S}_\sigma(\widehat{\sigma}^2_i|\sigma^2_p(R_i),\kappa_p=3).
\end{equation}
for which the parent line-of-sight velocity distributions $\mathcal{F}$ are assumed to be Gaussian.
In practice we use the Pearson family of distributions to generate the sampling distributions $\mathcal{S}$ for the joint analysis and refer the reader to \cite{me} for a visual and mathematical guide to these distributions and a more detailed outline of the numerical procedure.

\subsection{Monte-Carlo Markov Chain Methodology}
An efficient means of deriving posterior distributions for a highly dimensional data set $p$ is to employ a random walk Monte-Carlo Markov Chain (MCMC) that moves through the parameter space by proposing new positions $p^{\prime}$ in the chain according to the current location $p$ and accepting these positions according to the ratio of likelihoods with the Metropolis-Hastings \citep{metro} algorithm. An acceptance ratio of $0.2$ to $0.4$ strikes a balance between exploration and precision suitable for higher dimensional parameter spaces. We adopt a multivariate Gaussian proposal density $Q(p^{\prime}|p)$ wherein the covariance matrix is updated periodically from previous entries in the chain to optimise mixing of correlated parameters. A diagonal covariance matrix is used as an initial estimate where entries are determined by fixing all but one parameters and adjusting the effective univariate Gaussians width until an acceptance of $\approx 50\%$ is achieved. As the MCMC is maximally efficient when the proposal density matches the posterior distributions the following transformations \citep{charbonnier2011} and prior ranges are imposed on the anisotropy and density parameters,
\begin{eqnarray}
-\log_{10}[1-A]&:&[-1,1] \nonumber\\
\log_{10}B&:&[1,4]\nonumber\\
\log_{10}\rho_{-2}&:&[-8,5]\\
\log_{10}r_{\text{-2}}&:&[1,6]\nonumber \\
\log_{10}\alpha&:&[-1.3,1.3]\nonumber
\end{eqnarray}
where $A=\{\beta_0,\beta_{\infty},\beta'_0,\beta'_{\infty}\}$ and $B=\{r_{\beta},r'_{\beta}\}$. We run four MCMC chains for each analysis simultaneously from different starting positions and stop when the four MCMC chains satisfy the Gelmans-Rubins convergence criterion $R<1.03$. We also check \citep{charbonnier2011} that the parameter variances across the chains are smaller than $1\%$ of the mean of these variances. Inspecting the autocorrelation function of each chain we thin them out accordingly after removing the initial burn-in period. 

From the processed chains we are able to construct probability distributions and thus confidence intervals for a number of interesting physical quantities such as the enclosed mass and the logarithmic mass slope $\Gamma = \rm{d}\ln M/\rm{d}\ln r$. To do this a chain is created for each in a range of radii by evaluating these quantities from the elements in the parameter chain. Assuming flat priors on the derived quantities the median values then correspond to best estimates and one may use fractiles to determine the central confidence intervals, i.e if $95\%$ of $\Gamma(200\rm{pc})$ values in the chain lie between $\Gamma_u$ and $\Gamma_d$ then these become the $95\%$ confidence intervals at this radius.

\subsection{Test Data}

Though the statistical method presented above has been tested \citep{me} for constant anisotropy, the addition of four parameters to the generalised case adds an additional strain to the MCMC analysis that warrants further testing. Simulated data was generated as per \cite{me} from a parent distribution\footnote{Data was generated from the Gaussian Superposition family of distributions in appendix B of \cite{me} so as to again test the assumption of Pearson generated sampling distributions} with dispersion and kurtosis profiles corresponding to parameter set $p_{\rm{in}}$ which we aim to recover with the MCMC. As expected the additional parameters increased the number of iterations typically required to satisfy the convergence criterion and chains with between and $10^5$ and $10^6$ entries were thinned to one in every hundred for satisfactory autocorrelation values and independence. This could be improved in future analysis by adopting a more flexible proposal density that can identify the different correlations between parameters in different parts of the parameter space.

\begin{figure}
	\centering
		\includegraphics[width=9.0cm]{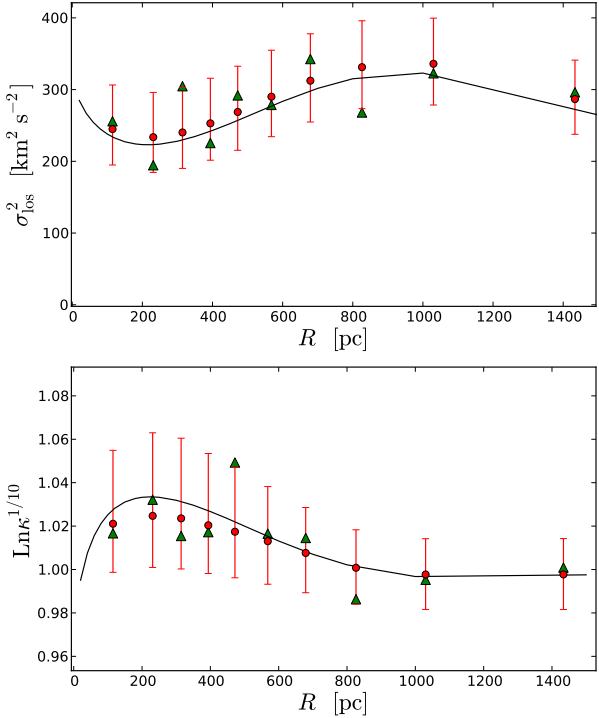}
        	\caption{Mock Fornax simulated data: Solid black curves indicate the the line of sight dispersion \eqref{LOSsecond} and kurtosis \eqref{LOSgenfour} used to generate the mock data set $p_{\rm{in}}$. Green triangles show the dispersion \eqref{sampdisp} and kurtosis of the mock data set. Red circles and error bars show the median and $95\%$ confidence interval for the reconstruction of the inpute parameters following the analysis in the text.}
	\label{simmoments}
\end{figure}

\begin{figure*}
	\centering
		\includegraphics[width=20cm]{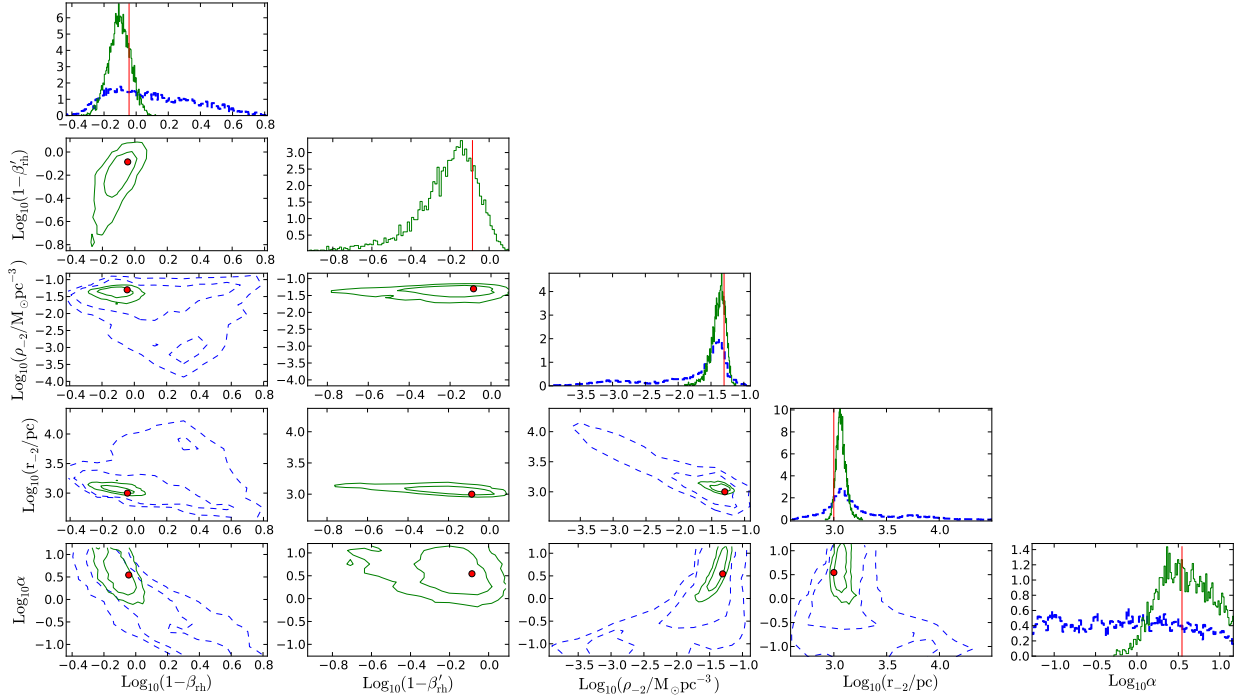}
        	\caption{MCMC reconstruction of mock Fornax data: The posterior distribution of dispersion-only analyses is marked in dashed blue while the dispersion-kurtosis analysis is solid-green. For brevity we collect the anisotropy parameters by plotting distributions of the anisotropy at the projected half light radius denoted $\beta_{\rm{rh}}$. Vertical red lines and red dots show the input parameters $p_{\rm{in}}$.}
	\label{simmcmc}
\end{figure*}

\begin{figure*}
	\centering
		\includegraphics[width=20cm]{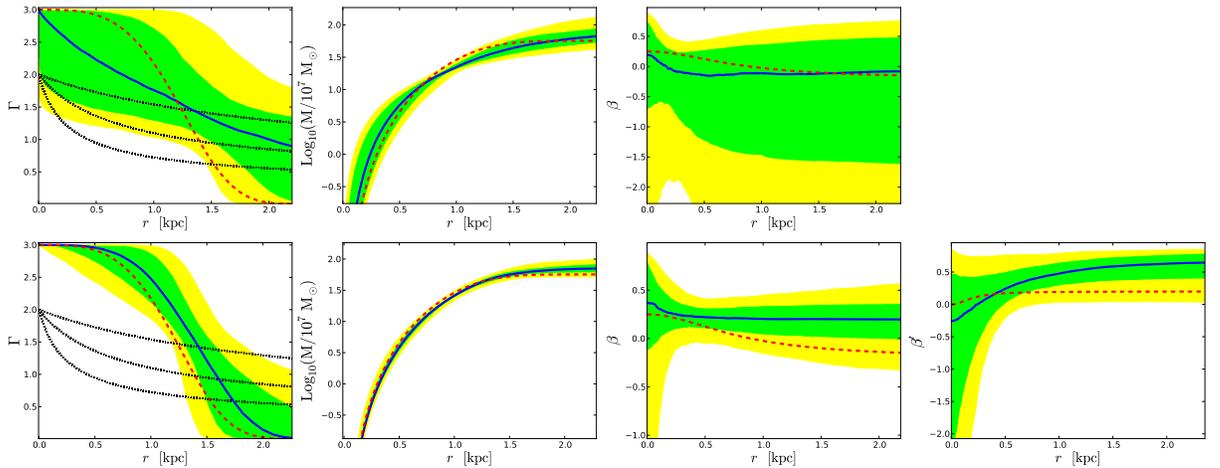}
        	\caption{Confidence intervals for mock Fornax data: Columns 1-3 show the median (solid blue line) and $67/95\%$ confidence intervals (respectively green and yellow shading) for the logarithmic mass slope, mass profile and anisotropy parameter as calculated from the derived posterior distributions output by the MCMC. Dashed red lines show true input parameters and in column 1 NFW profiles (dotted black) are plotted with three different scale radii, $r_s = (200,\; 600,\; 2000)\rm{pc}$, for reference. Column 4 is the fourth order anisotropy, relevant only to the dispersion-kurtosis analysis.}
	\label{simci}
\end{figure*}

 Fig. \ref{simmoments} shows the moment estimators of the simulated data and the moment profiles corresponding to  $p_{\rm{in}} = \{\beta_0=0.25,\beta_\infty=-0.2,r_\beta = 0.8\rm{kpc},\beta^{\prime}_{0}= 0.0,\beta^{\prime}_{\infty}=0.2,r^{\prime}_{\beta}=0.2\rm{kpc},\rho_{-2}=0.05\rm{M}_{\odot}\rm{pc}^{-3},r_{-2} = 1\rm{kpc},\alpha=3.5\}$ in the Jeans equations that is used to generate it. The error bars and central values show the confidence intervals and bias from the sampling distributions $\mathcal{S}(p_{\rm{in}})$. Figs. \ref{simmcmc} and \ref{simci} show that the MCMC was able to reproduce the input parameters $p_{\rm{in}}$ within $95\%$ confidence intervals for a mock Fornax data set of two thousand stars, a plummer profile with a projected half-light radius of $R_{1/2} = 575\rm{pc}$ and normally distributed experimental errors of $22\%$ relative to the intrinsic dispersion equivalent to a few $\rm{kms}^{-1}$. The input parameters were chosen such that the density had an extended core and the dispersion measurements are approximately constant at all radii as shown in Fig. \ref{simmoments} with the curved features masked by the natural scatter. Figs. \ref{simmcmc} and \ref{simci} show that the dispersion-only fit is not able to distinguish between radial and tangential anisotropy, nor is it able to tell the difference between NFW-like and cored halos. 

With the addition of the kurtosis measurement this degeneracy is broken with the dotted NFW profiles disfavoured to high significance and the cored nature of the simulated halo being detected.  This shows that in some circumstances it is possible with generalised as well as constant anisotropy to break the degeneracy in the traditional analysis. This result is of course aided by the chosen set of parameters with a kurtosis profile that has a distinct leptokurtic feature around 300pc that disfavours platykurtic ($\kappa^{\prime}<\ln(3)^{1/10} \approx 1.01$) points in that region. With a perfect assessment of the tracer density $\nu(r)$ and exact parametric forms for the anisotropy and density the simulation is always likely to be optimistic in its assessment of the confidence intervals but serves as a proof of concept and a crucial test of the MCMC's reliability.

\section{Application to Fornax and Sculptor Dwarf Spheroidals}
\subsection{Data}
Stellar position and velocity data for Fornax and Sculptor is taken from the Magellan survey \citep{walkdat} and after using the spectral metallicity data provided in \cite{walkermetal} to remove potential interlopers with probability of membership less than 0.95 one retains a sample of 2304 and 1351 stars for Fornax and Sculptor respectively. To ensure that there are at least two hundred stars in each bin which is a minimum for a meaningful measurement of the fourth moment \citep{Amorisco}, the Fornax data is split into nine radial annuli of 230 stars plus an outer annulus of 234 stars and in the same manner the Sculptor data is split into five annuli of 225 plus one of 226. To construct the sampling distributions (see fig. 7 and surrounding text in \cite{me}) for Fornax and Sculptor (denoted with superscripts f and s respectively) we generated bootstrap sample sizes from a Pearson distribution of $N^f =230$ and $N^s=225$ stars assuming scale dispersions of $\sigma^{f}_{s} = 10\rm{kms}^{-1}$ and $\sigma^{s}_{s} = 9\rm{kms}^{-1}$ and adding experimental errors \citep{Amorisco} normally distributed with widths $\delta^{f} = 0.22\sigma^{f}_{s}$ and $\delta^{s} = 0.325\sigma^{s}_{s}$. Projected stellar radii are calculated assuming galactic centres presented in \cite{mateo} and converted to parsecs with distance measurements from that paper. The half-light radius used in the Plummer model fit to the stellar density is taken from \cite{irwin} and for clarity are shown in Table \ref{galparams} with the aforementioned distances and central coordinates. 
\begin{table}
\caption{Galaxy Parameters: Central coordinates (J2000), distances and half-light radii}
\centering
\begin{tabular}{c c c c c}
\hline
Galaxy & $\alpha_c$ & $\delta_c$ & $d\;[\rm{kpc}]$ & $R_{1/2}\;[\rm{kpc}]$ \\
\hline
Fornax & $02^h39^m59^s$ & $-34^{o} 27.0^{\prime}$ & 138 & 0.67 \\
Sculptor  & $01^h00^m09^s$ & $-33^{o} 42.5^{\prime}$ & 79 & 0.26  \\
\hline
\end{tabular}
\label{galparams}
\end{table}      

\subsection{MCMC Results}

\subsubsection{Fornax}
%\begin{figure*}
	%\centering
	%	\includegraphics[width=17cm]{fornaxmcmc.eps}
        %	\caption{MCMC ouput for Fornax: Key as per Fig. \ref{simmcmc} but without red vertical lines and dots in the absence of known input parameters.}
%	\label{fornaxmcmc}
%\end{figure*}

\begin{figure*}
	\centering
		\includegraphics[width=20cm]{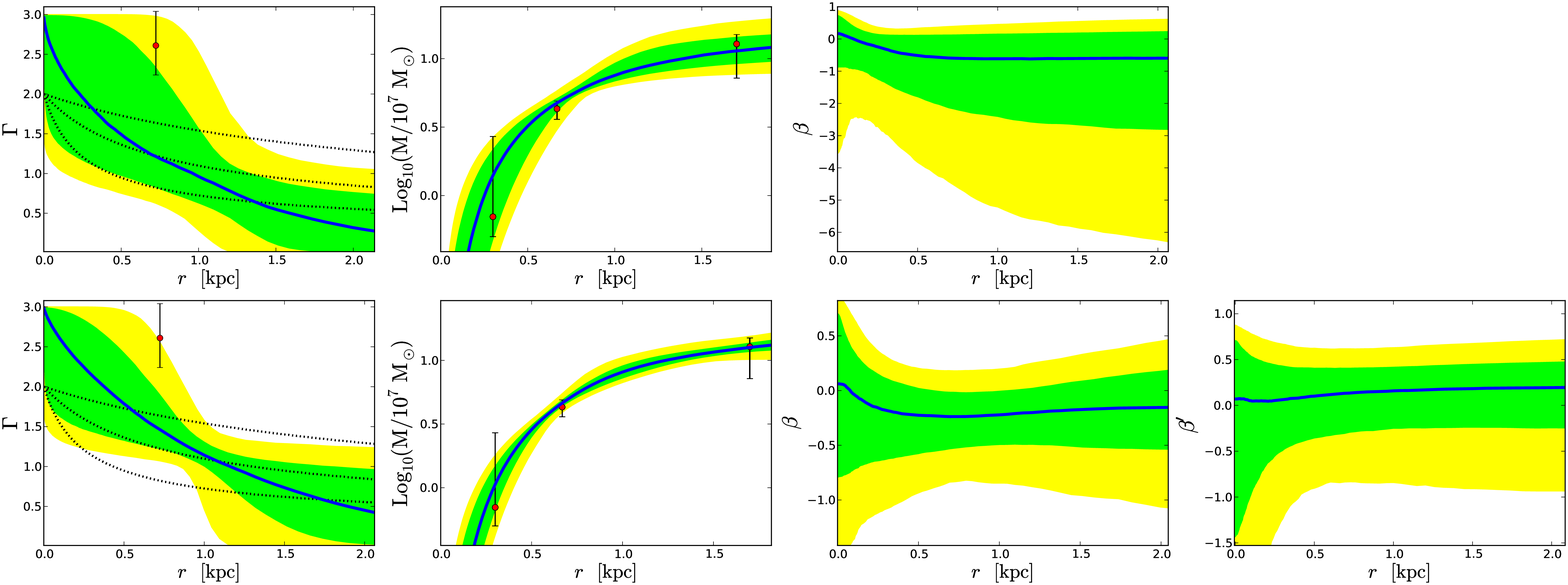}
       	\caption{Confidence intervals for Fornax: Key as per figure \ref{simci} but without red dashed lines to indicate the input parameters. In column 1 the data point and error bar corresponds to the mass slope and error derived for Fornax in \protect \cite{penarrubia} and for its radius we use the average of the two sub component half light radii. In column 2 the data points show the confidence intervals derived from the dispersion-only Jeans analysis in \protect \cite{universal} of the same data sets used here for $M(300\rm{pc})$, $M(R_{1/2})$ and $M(R_l)$ where $R_l$ is the projected radius of the last star.}

	\label{fornaxci}
\end{figure*}

 Considering first the dispersion-only analysis% denoted by blue dashed histograms and contours in Fig. \ref{fornaxmcmc}% and 
 in the upper panel of Fig. \ref{fornaxci} we see that as expected the anisotropy is almost completely obscured by the traditional degeneracy of the flat dispersion measurement with the $95\%$ confidence intervals at best enclosing $\beta(r)$ between moderately radial and strongly tangential orbits and placing even less constraint at the galactic centre and larger distances. The mass contraints for the dispersion-only analysis are similar to the displayed error bars for the dispersion-only analysis of the same data in \cite{universal} which uses more spatial bins, a more flexible Zhao \citep{Zhao} density profile and assumes a Gaussian distribution for $\mathcal{S}_\sigma$ in the likelihood. As shown in fig. 7 of \cite{me} the latter assumption is good provided the kurtosis is not more than mildly leptokurtic as is the case in Fornax. The deprojected radius $r^{\star}_{-3}$ at which the logarithmic density slope of the stellar density profile $\gamma_\star = -3$ and where the mass is robust to the anisotropy \citep{Wolf}, is related to the projected half-light radius via $r^{\star}_{-3} \approx 1.23 R_{1/2}$ for a Plummer profile and we see the characteristic pinch in the mass at $r^{\star}_{-3} = 822$pc. The mass slope confidence intervals in the upper left panel show that the dispersion-only analysis is unable to distinguish between the cusped NFW profiles in black and cored profiles that preserve the constant density with $\Gamma = 3$ out to around 750pc, at high significance though the NFW profile is mildly favoured.

Adding the kurtosis to the analysis % with solid green histograms and contours in Fig. \ref{fornaxmcmc} and%
in the bottom panels of Fig. \ref{fornaxci} we note first the dramatic improvement in precision of the anisotropy parameter $\beta$ at the half-light radius where there is maximal information from the data in the kurtosis measurement which is sensitive to the anisotropy \citep{Lokas05}. Uncertainty in the anisotropy at the centre of the galaxy is large which hopefully reflects that with little reference to the data at very small radii the freedom in the anisotropy profile removes any spurious effects from the assumption of a Plummer profile that limits the central behaviour of the dispersion and fourth moment. In the fourth column the anisotropy between fourth order moments $\beta^{\prime}$, constrained only by the kurtosis measurement, shows the new fourth order degeneracy as less affecting than the traditional degeneracy in $\beta$ in the plot to the left. For Fornax however it is not possible to distinguish between radially or tangentially biased kurtosis measurements. This could explain why only moderate improvement is made on the mass and mass slope constraints from the top to bottom panels in the two columns on the left in comparison to the simulated data. Indeed the primary effect on the Fornax MCMC of including the kurtosis is to squeeze the confidence intervals with little change to the best estimates. 

In summary, for Fornax cores and cusps are still not distinguished at high significance but whilst the NFW profile is still in excellent agreement, cored profiles as extended as the one obtained in \citep{penarrubia} are in slight tension with the data which favours a density profile with an inner slope somewhere between the two. This result is also contrary to predictions from the virial theorem \citep{amorfornax} and orbit superposition methods \citep{Jardel} though the latter only provides likelihoods for NFW and a small set of cored profiles. 

One interesting improvement in the joint analysis is in the precision of the scale radius measurement which shows as a pinch in the mass slope confidence intervals. For cored profiles in particular having two references to the data in the joint analysis one envisages that this helps to distinguish this natural turning point in the moment profiles from the anisotropy. As the kurtosis is effectively independent of the scale density the degeneracy between the scale density and radius inherent to the overall scaling of the dispersion is partially broken. 
\subsubsection{Sculptor}
  
%\begin{figure*}
%	\centering
%		\includegraphics[width=17cm]{sculptormcmc.eps}
%        	\caption{MCMC ouput for Sculptor: Key as per Fig. \ref{fornaxmcmc}}
%	\label{sculptormcmc}
%\end{figure*}

\begin{figure*}
	\centering
		\includegraphics[width=20cm]{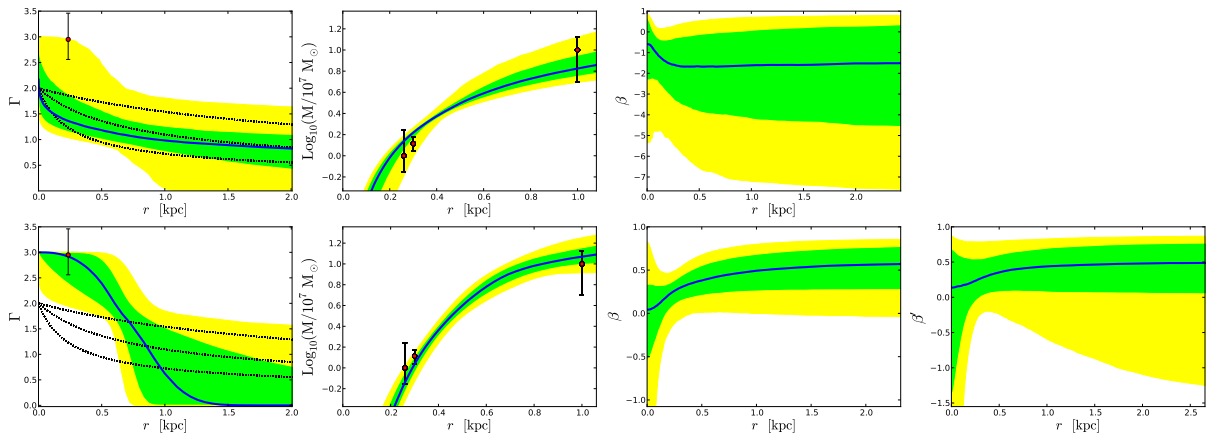}
        	\caption{Confidence intervals for Sculptor: Key as per Fig. \ref{fornaxci}.}
	\label{sculptorci}
\end{figure*}

The dispersion-only analysis of Sculptor shown% indicated by blue dashed lines in Fig. \ref{sculptormcmc} and
in the top panels of \ref{sculptorci} also shows a large degeneracy in the anisotropy parameter but here with a clear bias towards tangential velocity anisotropies.  Both NFW profiles (dotted black) and profiles with cores out to the half light radius are both withing the $95\%$ confidence interval.  The $67\%$ confidence interval slightly favours the steeper NFW profile, in tension with the data point obtained in \cite{penarrubia} using their multiple stellar population method.

Caution is urged in placing too much emphasis on this second order result due to our choosing only 6 radial bins in order to later directly compare our second and fourth order analyses (as mentioned earlier, in order to reliably fit the kurtosis we require bins containing a relatively large number of stars).  Firstly, this relatively small number of bins compared to studies in the literature that adopt as many as $\sqrt{N}$ bins for the variance is not an optimal balance of spatial and statistical precision. With little reference to the outermost stars this choice places a strong emphasis on the outermost data point which suggests a rising velocity dispersion whilst including more bins, as in \cite{universal}, indicates that the dispersion then falls again at larger radii making it more compatible with cored fits. As such we take this result with a grain of salt and place more emphasis on the $95\%$ confidence intervals that show the degeneracy that has been found in the literature.  This is the reason for any small difference between the results of our second order analysis and other in the literature.

The inclusion of the kurtosis for sculptor is rather more dramatic than for Fornax. Here we see that though the anisotropy at the galactic centre is obscured the joint analysis makes a clear prediction for radial anisotropy. This is partly mirrored by the anisotropy between fourth order moments $\beta^{\prime}$ though at large radii the significance is smaller owing to the fourth order degeneracy. More interesting still, the cored solution, in excellent agreement with the prediction from multiple stellar populations, is favoured over the NFW one to high significance up to and beyond the half-light radius breaking the degeneracy. This is also manifest in the mass estimate where systematically lower masses are predicted at small radii which would reduce the astrophysical J-factor over small integration angles. On face value then we validate the prediction of a cored density profile in studies of multiple populations \citep{penarrubia, amoriscomulti} and from the virial theorem \citep{evansvirial}. 

\begin{figure}
	\centering
		\includegraphics[width=9cm]{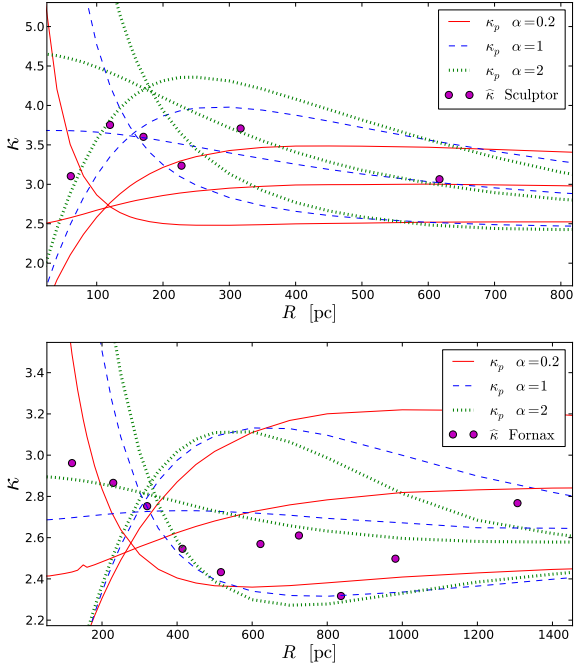}
        	\caption{Generic kurtosis profiles and data: Kurtosis \eqref{sampkurt} is plotted for the ten radial bins of Fornax (lower panel) and 6 of Sculptor (upper panel). Against these we plot line of sight kurtosis profiles from the Jeans equations assuming Plummer profiles for each with half-light radii from Table 1 and Einasto scale radii of $r_{-2} =$700pc similar to the median scale radii $r_{-2} = $698 and 641pc returned by the fourth order MCMC for Fornax and Sculptor respectively. With no attempt made at fitting the data we show three constant anisotropy models with $\beta = (-0.6,0,0.4)$ and isotropy at fourth order $\beta^{\prime}=0$ for each adopted shape parameter $\alpha$ that is distinguished by colour and line style. A simple means to distinguish between anisotropy is to note that tangential curves diverge at the centre, isotropic curves tend to a finite constant value and radially anisotropic curves vanish at the origin. Curves in both plots have the same DM density and anisotropy parameters with differences arising because of the change in the adopted half-light radius.}
	\label{kurtosisvary}
\end{figure}

\subsection{Why is the Fourth Order Method Better for Sculptor and not for Fornax?}

The results for Sculptor and the simulated data imply that the kurtosis measurements are key to distinguishing between the scale radius $r_{-2}$ and shape parameter $\alpha$, a degeneracy that normally exists with a flat velocity dispersion profile, however there is no corresponding significant improvement for Fornax. Inspection of the kurtosis data in Fig. \ref{kurtosisvary} shows that while Fornax has an approximately flat series of kurtosis measurements that dips below the Gaussian value of $\kappa = 3$, the sculptor data exhibits a group of leptokurtic points ($\kappa > 3$) in the dense stellar region before dipping again to Gaussianity at the end confirming the results in \cite{Amorisco} and \cite{Breddels}. With the assumption of mild anisotropy we see that the kurtosis profiles converge at a point below the half-light radius. The location of this convergence point on the kurtosis axes displays a dependence on the shape parameter $\alpha$, an increase in which shifts it upwards.

Comparing solutions with identical DM density profiles in the upper and lower panels of Fig. \ref{kurtosisvary} we see that by extending the half-light radius in the Jeans equation the scale of the shift is reduced and the convergence points are less separated on the kurtosis axis making it harder to distinguish between them. For Fornax where the median DM scale radius returned by the MCMC $\log_{10}\;r_{-2} = 2.81^{+0.11(0.94)}_{-0.08(0.20)}$ is similar to the projected half-light radius $r_{-2}/R_{1/2} \approx 1$ we see that the shift is reduced compared to Sculptor which with a similar DM scale radius $\log_{10}\;r_{-2} = 2.81^{+0.17(1.28)}_{-0.07(0.13)}$ has a more embedded stellar population $r_{-2}/R_{1/2} \approx 2.5$. The NFW-like density profile with $\alpha = 0.2$ is least sensitive to this effect and a very large scale radius(and thus diffuse halo to fit the dispersion data) is required to shift the convergence point towards the Gaussian value $\kappa = 3$. For this reason it is very difficult to generate leptokurtic points such as those in the Sculptor data set with $\widehat{\kappa}>3.5$ that lie near the convergence radius. This is exacerbated  by the fact that fat tailed platykurtic distributions are less prone to large statistical fluctuations which reduces the likelihood of points in the outer tails.

In conclusion there appears to be a radius where the kurtosis is invariant to the anisotropy which for NFW-like profiles falls well below $\kappa = 3$. While radial anisotropy shifts the curve upwards towards the Sculptor data points at larger radii, the kurtosis between 100 and 150pc is stuck and cannot describe the data points with $\kappa>3.5$. For this reason not only is radial anisotropy preferred in Sculptor but the MCMC is able to fit cored profiles with higher likelihood than steeper ones regardless of the anisotropy. In Fornax the data points have kurtosis $\kappa<3$ and are easier to describe with NFW-like profiles. The extended half-light radius also reduces the ability to distinguish between DM density profiles at the point where the kurtosis is invariant to the anisotropy. If these empirical results stand up to a wider range choices for the anisotropy parameters or a variation in the adopted form of the stellar density profile then Sculptor's leptokurtic data points could be a smoking gun for a cored density profile and we are motivated (Richardson $\&$ Fairbairn in preparation) to confirm these findings with analytic reasoning.

% Indeed we found that if one chooses more extreme values $\beta >> \beta^{\prime}$ then it is possible to shift the convergence point to the leptokurtic region for NFW-like profiles with large scale radii. This result that follows from a dominant second term in \eqref{LOSgenfour} suggests a fine tuned situation wherein the radial velocity distribution 
                    
\subsection{Key Assumptions and Limitations}

In this subsection we address the potential impact of the simplifying assumptions inherent to the Jeans analysis used above as well as astrophysical processes and phenomena that, for simplicity, were omitted. 

The assumption of spherical symmetry, necessary to close the set of moment equations, has been shown \citep{evanscentre} to have a significant impact on the dynamics at the Galactic centre when coupled with restrictions on the stellar and dark matter logarithmic density slopes. Though concern is raised that this artifact of the assumptions could lead to spurious results as discussed in Section 3.1 we take comfort from the fact that the uncertainty in the anisotropy parameter derived from the MCMC is large at the centre, thus minimising its influence with the anisotropy in the densest stellar regions free to vary independently from these constraints. For the joint analysis, we note that the confidence intervals narrow considerably at a finite radius near that of half light where there are most stars and thus information from the data and that for Sculptor, with the strongest constraints this effect is particularly pronounced. This suggests that adopting a generalised radial profile for the anisotropy is important to decouple this useful behaviour from the assumptions though to confirm these findings one should perform an analysis where both inner slopes have a greater degree of freedom.

Photometric measurements \citep{irwin} of Fornax and Sculptor indicate a departure from spherical symmetry with the ratio of major and minor axes $\approx 0.7$. By considering elliptical rather than circular annuli it has been shown \citep{penarrubia} that the mass slopes derived by studying multiple populations, where uncertainties in the half light radii are important, are sensitive with Sculptor dropping from $\Gamma = 2.95$ to $\Gamma = 2.4$. By contrast \cite{Amorisco} find that binned measurements of the sample moments are broadly similar in the minor and major axis directions and with more spatial points of reference the Jeans analysis should be more robust. 

The other major assumption inherent to the Jeans analysis is dynamic equilibrium though significant tidal disruption in the classical dwarfs has only been conclusively identified \citep{saggit} in Sagittarius. Though simulations \citep{tides} suggested that the Jeans analysis is robust to tidal forces with suitable interloper removal procedures, it has recently been suggested that measurements obtained with the mass slope method \citep{penarrubia} are subject to systematic uncertainties dependent on the unknown line of sight. One of the great powers of this method, however, is that it can be applied to small samples and an analysis of a larger group of data sets would reduce any systematic bias. 
            
Binary stars could also inflate the dispersions \citep{binaries} though the case has been put forward \citep{ufaintbin} that significant effects only occur for dispersions on the scale of the ultra faint dwarfs with $\sigma_p\leq 3\rm{kms}^{-1}$ significantly smaller than those of the classical dSphs considered herein with $\sigma_p \approx 10\rm{kms}^{-1}$. 

\section{Discussion}

In \cite{me} a Jeans analysis of the variance and kurtosis was devised that for the first time fully extended the framework from second order to fourth order without imposing restrictions on the anisotropy between fourth order moments which we introduced as $\beta^{\prime}$. 

With generic and uncorrelated parametric forms for the radial profiles of $\beta$ and $\beta^{\prime}$ we applied the analysis to kinematic data sets for the Milky Way dwarf spheroidals Fornax and Sculptor in the hope of breaking the mass-anisotropy degeneracy that plagues dispersion-only analyses of real data sets with flat dispersion curves. The assumption of spherical symmetry and a fixed parametrised form for the stellar density profile $\nu(r)$ has a significant impact upon the Jeans equations at the galactic centre and the freedom to divorce the central anisotropy from that at regions of high stellar density provides more confidence than previous joint analyses \citep{Lokas05} that have been limited to constant anisotropy. 

 Though the joint analysis of Fornax was able to substantially improve the precision in the anisotropy measurement, little improvement was made for the mass slope which could not distinguish between an NFW-like Einasto profile and one with an extended core at high significance. Slight tension was however found with large cores predicted in the literature from studies of multiple populations with the best fits being intermediates between these and the NFW-like models. By contrast the analysis of Sculptor favoured radial anisotropy and showed a clear preference for cored solutions in excellent agreement with the literature. At first glance it appears that the rather distinct kurtosis measurements for the two galaxies are key to interpreting these results and we demonstrate that the leptokurtic data points in Sculptor seem difficult to reproduce with an NFW-like profile when one assumes moderate anisotropy.Further investigation into this effect with a wider range of anisotropy and stellar density models is required to confirm this finding which we hope to present in the near future. As both galaxies display two very different anisotropy and density profiles it could be argued that astrophysical effects, more chaotic in nature, play a significant role in shaping the density profiles of dwarf spheroidal galaxies that could obscure systematic effects induced by warm or self interacting dark matter cosmologies for which a larger sample of galaxies is needed. 

In summary this method provides an analytical insight into the dynamics of spherical systems with a generality bettered only by numerical orbital superposition methods. It also complements the bold predictions from stellar sub component methods by testing the assumption of Gaussianity. With further investigation into the nature of the kurtosis profile we hope to provide simple analytic insights that, like the robust mass estimate at the half-light radii, could give simple justification to the interesting results of the MCMC analysis presented here and help to settle the cusp vs core debate.    

\section*{Acknowledgments}
TR thanks the KCL Graduate School for support.  MF is a grateful recipient of support from the STFC and enjoyed related discussions with Riccardo Catena and Roberto Trotta.

\bibliographystyle{mn2e2}
\bibliography{tranrep}

\begin{thebibliography}{54}
\expandafter\ifx\csname natexlab\endcsname\relax\def\natexlab#1{#1}\fi

\bibitem[{{Ackermann} {et~al}\mbox{.}(2011){Ackermann}, {Ajello}, {Albert},
  {Atwood}, {Baldini}, {Ballet}, {Barbiellini}, {Bastieri}, {Bechtol},
  {Bellazzini}, {Berenji}, {Blandford}, {Bloom}, {Bonamente}, {Borgland},
  {Bregeon}, {Brigida}, {Bruel}, {Buehler}, {Burnett}, {Buson}, {Caliandro},
  {Cameron}, {Ca{\~n}adas}, {Caraveo}, {Casandjian}, {Cecchi}, {Charles},
  {Chekhtman}, {Chiang}, {Ciprini}, {Claus}, {Cohen-Tanugi}, {Conrad},
  {Cutini}, {de Angelis}, {de Palma}, {Dermer}, {Digel}, {Do Couto E Silva},
  {Drell}, {Drlica-Wagner}, {Falletti}, {Favuzzi}, {Fegan}, {Ferrara},
  {Fukazawa}, {Funk}, {Fusco}, {Gargano}, {Gasparrini}, {Gehrels}, {Germani},
  {Giglietto}, {Giordano}, {Giroletti}, {Glanzman}, {Godfrey}, {Grenier},
  {Guiriec}, {Gustafsson}, {Hadasch}, {Hayashida}, {Hays}, {Hughes}, {Jeltema},
  {J{\'o}hannesson}, {Johnson}, {Johnson}, {Kamae}, {Katagiri}, {Kataoka},
  {Kn{\"o}dlseder}, {Kuss}, {Lande}, {Latronico}, {Lionetto}, {Llena Garde},
  {Longo}, {Loparco}, {Lott}, {Lovellette}, {Lubrano}, {Madejski}, {Mazziotta},
  {McEnery}, {Mehault}, {Michelson}, {Mitthumsiri}, {Mizuno}, {Monte},
  {Monzani}, {Morselli}, {Moskalenko}, {Murgia}, {Naumann-Godo}, {Norris},
  {Nuss}, {Ohsugi}, {Okumura}, {Omodei}, {Orlando}, {Ormes}, {Ozaki},
  {Paneque}, {Parent}, {Pesce-Rollins}, {Pierbattista}, {Piron}, {Pivato},
  {Porter}, {Profumo}, {Rain{\`o}}, {Razzano}, {Reimer}, {Reimer}, {Ritz},
  {Roth}, {Sadrozinski}, {Sbarra}, {Scargle}, {Schalk}, {Sgr{\`o}}, {Siskind},
  {Spandre}, {Spinelli}, {Strigari}, {Suson}, {Tajima}, {Takahashi}, {Tanaka},
  {Thayer}, {Thayer}, {Thompson}, {Tibaldo}, {Tinivella}, {Torres}, {Troja},
  {Uchiyama}, {Vandenbroucke}, {Vasileiou}, {Vianello}, {Vitale}, {Waite},
  {Wang}, {Winer}, {Wood}, {Wood}, {Yang}, {Zimmer}, {Kaplinghat}, \&
  {Martinez}}]{fermidwarf}
{Ackermann} M. {et~al.}, 2011, Physical Review Letters, 107, 241302

\bibitem[{{Agnello} \& {Evans}(2012)}]{evansvirial}
{Agnello} A., {Evans} N.~W., 2012, \apjl, 754, L39

\bibitem[{{Amorisco} {et~al}\mbox{.}(2012){Amorisco}, {Agnello}, \&
  {Evans}}]{amorfornax}
{Amorisco} N.~C., {Agnello} A., {Evans} N.~W., 2012, ArXiv e-prints

\bibitem[{{Amorisco} \& {Evans}(2012{\natexlab{a}})}]{amoriscomulti}
{Amorisco} N.~C., {Evans} N.~W., 2012{\natexlab{a}}, \mnras, 419, 184

\bibitem[{{Amorisco} \& {Evans}(2012{\natexlab{b}})}]{Amorisco}
{Amorisco} N.~C., {Evans} N.~W., 2012{\natexlab{b}}, \mnras, 424, 1899

\bibitem[{{An}(2011)}]{an11a}
{An} J.~H., 2011, \mnras, 413, 2554

\bibitem[{{Baes} \& {van Hese}(2007)}]{baes07}
{Baes} M., {van Hese} E., 2007, \aap, 471, 419

\bibitem[{{Binney} \& {Tremaine}(2008)}]{binney}
{Binney} J., {Tremaine} S., 2008, {Galactic Dynamics: Second Edition}.
  Princeton University Press

\bibitem[{{Boyarsky} {et~al}\mbox{.}(2009){Boyarsky}, {Lesgourgues},
  {Ruchayskiy}, \& {Viel}}]{warmoklarge}
{Boyarsky} A., {Lesgourgues} J., {Ruchayskiy} O., {Viel} M., 2009, Physical
  Review Letters, 102, 201304

\bibitem[{{Boylan-Kolchin} {et~al}\mbox{.}(2011){Boylan-Kolchin}, {Bullock}, \&
  {Kaplinghat}}]{toobigtofail}
{Boylan-Kolchin} M., {Bullock} J.~S., {Kaplinghat} M., 2011, \mnras, 415, L40

\bibitem[{{Boylan-Kolchin} {et~al}\mbox{.}(2010){Boylan-Kolchin}, {Springel},
  {White}, \& {Jenkins}}]{noplacelikehome}
{Boylan-Kolchin} M., {Springel} V., {White} S.~D.~M., {Jenkins} A., 2010,
  \mnras, 406, 896

\bibitem[{{Breddels} {et~al}\mbox{.}(2012){Breddels}, {Helmi}, {van den Bosch},
  {van de Ven}, \& {Battaglia}}]{Breddels}
{Breddels} M.~A., {Helmi} A., {van den Bosch} R.~C.~E., {van de Ven} G.,
  {Battaglia} G., 2012, ArXiv e-prints

\bibitem[{{Charbonnier} {et~al}\mbox{.}(2011){Charbonnier}, {Combet}, {Daniel},
  {Funk}, {Hinton}, {Maurin}, {Power}, {Read}, {Sarkar}, {Walker}, \&
  {Wilkinson}}]{charbonnier2011}
{Charbonnier} A. {et~al.}, 2011, \mnras, 418, 1526

\bibitem[{{Col{\'{\i}}n} {et~al}\mbox{.}(2000){Col{\'{\i}}n}, {Avila-Reese}, \&
  {Valenzuela}}]{warmlesssats}
{Col{\'{\i}}n} P., {Avila-Reese} V., {Valenzuela} O., 2000, \apj, 542, 622

\bibitem[{{Conn} {et~al}\mbox{.}(2013){Conn}, {Lewis}, {Ibata}, {Parker},
  {Zucker}, {McConnachie}, {Martin}, {Valls-Gabaud}, {Tanvir}, {Irwin},
  {Ferguson}, \& {Chapman}}]{satphase}
{Conn} A.~R. {et~al.}, 2013, \apj, 766, 120

\bibitem[{{Dejonghe}(1986)}]{dejonghe86}
{Dejonghe} H., 1986, \physrep, 133, 217

\bibitem[{{Einasto} \& {Haud}(1989)}]{einasto}
{Einasto} J., {Haud} U., 1989, \aap, 223, 89

\bibitem[{{Evans} {et~al}\mbox{.}(2009){Evans}, {An}, \&
  {Walker}}]{evanscentre}
{Evans} N.~W., {An} J., {Walker} M.~G., 2009, \mnras, 393, L50

\bibitem[{{Gentile} {et~al}\mbox{.}(2004){Gentile}, {Salucci}, {Klein},
  {Vergani}, \& {Kalberla}}]{spiralcores}
{Gentile} G., {Salucci} P., {Klein} U., {Vergani} D., {Kalberla} P., 2004,
  \mnras, 351, 903

\bibitem[{{Governato} {et~al}\mbox{.}(2012){Governato}, {Zolotov}, {Pontzen},
  {Christensen}, {Oh}, {Brooks}, {Quinn}, {Shen}, \& {Wadsley}}]{nomorecusp}
{Governato} F. {et~al.}, 2012, \mnras, 422, 1231

\bibitem[{{Hargreaves} {et~al}\mbox{.}(1996){Hargreaves}, {Gilmore}, \&
  {Annan}}]{binaries}
{Hargreaves} J.~C., {Gilmore} G., {Annan} J.~D., 1996, \mnras, 279, 108

\bibitem[{{Irwin} \& {Hatzidimitriou}(1995)}]{irwin}
{Irwin} M., {Hatzidimitriou} D., 1995, \mnras, 277, 1354

\bibitem[{{Jardel} \& {Gebhardt}(2012)}]{Jardel}
{Jardel} J.~R., {Gebhardt} K., 2012, \apj, 746, 89

\bibitem[{{Klimentowski} {et~al}\mbox{.}(2007){Klimentowski}, {{\L}okas},
  {Kazantzidis}, {Prada}, {Mayer}, \& {Mamon}}]{tides}
{Klimentowski} J., {{\L}okas} E.~L., {Kazantzidis} S., {Prada} F., {Mayer} L.,
  {Mamon} G.~A., 2007, \mnras, 378, 353

\bibitem[{{Klypin} {et~al}\mbox{.}(1999){Klypin}, {Kravtsov}, {Valenzuela}, \&
  {Prada}}]{misssats}
{Klypin} A., {Kravtsov} A.~V., {Valenzuela} O., {Prada} F., 1999, \apj, 522, 82

\bibitem[{{Klypin} {et~al}\mbox{.}(2011){Klypin}, {Trujillo-Gomez}, \&
  {Primack}}]{bolshoisim}
{Klypin} A.~A., {Trujillo-Gomez} S., {Primack} J., 2011, \apj, 740, 102

\bibitem[{{Kroupa} {et~al}\mbox{.}(2005){Kroupa}, {Theis}, \&
  {Boily}}]{mwpancake}
{Kroupa} P., {Theis} C., {Boily} C.~M., 2005, \aap, 431, 517

\bibitem[{{{\L}okas}(2002)}]{Lokas02}
{{\L}okas} E.~L., 2002, \mnras, 333, 697

\bibitem[{{{\L}okas} {et~al}\mbox{.}(2005){{\L}okas}, {Mamon}, \&
  {Prada}}]{Lokas05}
{{\L}okas} E.~L., {Mamon} G.~A., {Prada} F., 2005, \mnras, 363, 918

\bibitem[{{Lovell} {et~al}\mbox{.}(2012){Lovell}, {Eke}, {Frenk}, {Gao},
  {Jenkins}, {Theuns}, {Wang}, {White}, {Boyarsky}, \& {Ruchayskiy}}]{warmsize}
{Lovell} M.~R. {et~al.}, 2012, \mnras, 420, 2318

\bibitem[{{Majewski} {et~al}\mbox{.}(2003){Majewski}, {Skrutskie}, {Weinberg},
  \& {Ostheimer}}]{saggit}
{Majewski} S.~R., {Skrutskie} M.~F., {Weinberg} M.~D., {Ostheimer} J.~C., 2003,
  \apj, 599, 1082

\bibitem[{{Mateo}(1998)}]{mateo}
{Mateo} M.~L., 1998, \araa, 36, 435

\bibitem[{{McConnachie} \& {C{\^o}t{\'e}}(2010)}]{ufaintbin}
{McConnachie} A.~W., {C{\^o}t{\'e}} P., 2010, \apjl, 722, L209

\bibitem[{{Merrifield} \& {Kent}(1990)}]{merry}
{Merrifield} M.~R., {Kent} S.~M., 1990, \aj, 99, 1548

\bibitem[{{Merritt}(1987)}]{merritt87}
{Merritt} D., 1987, \apj, 313, 121

\bibitem[{{Metropolis} {et~al}\mbox{.}(1953){Metropolis}, {Rosenbluth},
  {Rosenbluth}, {Teller}, \& {Teller}}]{metro}
{Metropolis} N., {Rosenbluth} A.~W., {Rosenbluth} M.~N., {Teller} A.~H.,
  {Teller} E., 1953, \jcp, 21, 1087

\bibitem[{Moore {et~al}\mbox{.}(1999)Moore, Ghigna, Governato, Lake, Quinn,
  {et~al.}}]{wherearethey}
Moore B., Ghigna S., Governato F., Lake G., Quinn T.~R., {et~al.}, 1999,
  Astrophys.J., 524, L19

\bibitem[{{Navarro} {et~al}\mbox{.}(1996){Navarro}, {Frenk}, \&
  {White}}]{NFWhalos}
{Navarro} J.~F., {Frenk} C.~S., {White} S.~D.~M., 1996, \apj, 462, 563

\bibitem[{{Read} \& {Gilmore}(2005)}]{astrocorefeedback}
{Read} J.~I., {Gilmore} G., 2005, \mnras, 356, 107

\bibitem[{{Richardson} \& {Fairbairn}(2012)}]{me}
{Richardson} T., {Fairbairn} M., 2012, ArXiv e-prints

\bibitem[{{Schwarzschild}(1979)}]{Schwarzschild}
{Schwarzschild} M., 1979, \apj, 232, 236

\bibitem[{{Spergel} \& {Steinhardt}(2000)}]{firstSI}
{Spergel} D.~N., {Steinhardt} P.~J., 2000, Physical Review Letters, 84, 3760

\bibitem[{{Springel} {et~al}\mbox{.}(2008){Springel}, {Wang}, {Vogelsberger},
  {Ludlow}, {Jenkins}, {Helmi}, {Navarro}, {Frenk}, \& {White}}]{aquarious}
{Springel} V. {et~al.}, 2008, \mnras, 391, 1685

\bibitem[{{Springel} {et~al}\mbox{.}(2005){Springel}, {White}, {Jenkins},
  {Frenk}, {Yoshida}, {Gao}, {Navarro}, {Thacker}, {Croton}, {Helly},
  {Peacock}, {Cole}, {Thomas}, {Couchman}, {Evrard}, {Colberg}, \&
  {Pearce}}]{NFWls}
{Springel} V. {et~al.}, 2005, \nat, 435, 629

\bibitem[{{Strigari} \& {Wechsler}(2012)}]{milkywayrare}
{Strigari} L.~E., {Wechsler} R.~H., 2012, \apj, 749, 75

\bibitem[{{Tegmark} {et~al}\mbox{.}(2004){Tegmark}, {Strauss}, {Blanton},
  {Abazajian}, {Dodelson}, {Sandvik}, {Wang}, {Weinberg}, {Zehavi}, {Bahcall},
  {Hoyle}, {Schlegel}, {Scoccimarro}, {Vogeley}, {Berlind}, {Budavari},
  {Connolly}, {Eisenstein}, {Finkbeiner}, {Frieman}, {Gunn}, {Hui}, {Jain},
  {Johnston}, {Kent}, {Lin}, {Nakajima}, {Nichol}, {Ostriker}, {Pope},
  {Scranton}, {Seljak}, {Sheth}, {Stebbins}, {Szalay}, {Szapudi}, {Xu},
  {Annis}, {Brinkmann}, {Burles}, {Castander}, {Csabai}, {Loveday}, {Doi},
  {Fukugita}, {Gillespie}, {Hennessy}, {Hogg}, {Ivezi{\'c}}, {Knapp}, {Lamb},
  {Lee}, {Lupton}, {McKay}, {Kunszt}, {Munn}, {O'Connell}, {Peoples}, {Pier},
  {Richmond}, {Rockosi}, {Schneider}, {Stoughton}, {Tucker}, {vanden Berk},
  {Yanny}, \& {York}}]{SDSS}
{Tegmark} M. {et~al.}, 2004, \prd, 69, 103501

\bibitem[{{Tonini} {et~al}\mbox{.}(2006){Tonini}, {Lapi}, \&
  {Salucci}}]{astrocore}
{Tonini} C., {Lapi} A., {Salucci} P., 2006, \apj, 649, 591

\bibitem[{{Walker} {et~al}\mbox{.}(2009){Walker}, {Mateo}, \&
  {Olszewski}}]{walkdat}
{Walker} M.~G., {Mateo} M., {Olszewski} E.~W., 2009, \aj, 137, 3100

\bibitem[{Walker {et~al}\mbox{.}(2009)Walker, Mateo, Olszewski, Peñarrubia,
  Evans, \& Gilmore}]{universal}
Walker M.~G., Mateo M., Olszewski E.~W., Peñarrubia J., Evans N.~W., Gilmore
  G., 2009, The Astrophysical Journal, 704, 1274

\bibitem[{{Walker} {et~al}\mbox{.}(2009){Walker}, {Mateo}, {Olszewski}, {Sen},
  \& {Woodroofe}}]{walkermetal}
{Walker} M.~G., {Mateo} M., {Olszewski} E.~W., {Sen} B., {Woodroofe} M., 2009,
  \aj, 137, 3109

\bibitem[{Walker \& Penarrubia(2011)}]{penarrubia}
Walker M.~G., Penarrubia J., 2011, Astrophys.J., 742, 20

\bibitem[{{Wolf} {et~al}\mbox{.}(2010){Wolf}, {Martinez}, {Bullock},
  {Kaplinghat}, {Geha}, {Mu{\~n}oz}, {Simon}, \& {Avedo}}]{Wolf}
{Wolf} J., {Martinez} G.~D., {Bullock} J.~S., {Kaplinghat} M., {Geha} M.,
  {Mu{\~n}oz} R.~R., {Simon} J.~D., {Avedo} F.~F., 2010, \mnras, 406, 1220

\bibitem[{{Zavala} {et~al}\mbox{.}(2013){Zavala}, {Vogelsberger}, \&
  {Walker}}]{walkSI}
{Zavala} J., {Vogelsberger} M., {Walker} M.~G., 2013, \mnras, 431, L20

\bibitem[{{Zhao}(1996)}]{Zhao}
{Zhao} H., 1996, \mnras, 278, 488

\end{thebibliography}

\end{document}